\documentclass{article}

\usepackage[preprint]{neurips_2024}

\usepackage[utf8]{inputenc} 
\usepackage[T1]{fontenc}    
\usepackage{hyperref}       
\usepackage{url}            
\usepackage{booktabs}       
\usepackage{amsfonts}       
\usepackage{nicefrac}       
\usepackage{microtype}      
\usepackage{xcolor}         

\usepackage[ruled, lined, linesnumbered, commentsnumbered, longend]{algorithm2e}

\usepackage{lineno}
\linenumbers
\nolinenumbers

\usepackage{amsmath}
\usepackage{amsmath}
\usepackage{multirow}
\usepackage{subcaption} 

\usepackage{graphicx, verbatim, url, pifont}
\usepackage{subfloat}
\usepackage{natbib}
\setcitestyle{numbers,open={[},close={]}} 

\title{PerLLM: Personalized Inference Scheduling with Edge-Cloud Collaboration for Diverse LLM Services}

\author{%
  Zheming Yang\textsuperscript{\rm 1,2}, Yuanhao Yang\textsuperscript{\rm 1,2}, Chang Zhao\textsuperscript{\rm 1,2}, Qi Guo\textsuperscript{\rm 1,2}, Wenkai He\textsuperscript{\rm 1,2}, Wen Ji\textsuperscript{\rm 1} \\
  \textsuperscript{\rm 1}Institute of Computing Technology, Chinese Academy of Sciences\\
\textsuperscript{\rm 2}University of Chinese Academy of Sciences\\
 \texttt{yangzheming19b@ict.ac.cn} \\
}

\begin{document}

\maketitle

\begin{abstract}
With the rapid growth in the number of large language model (LLM) users, it is difficult for bandwidth-constrained cloud servers to simultaneously process massive LLM services in real-time. Recently, edge-cloud infrastructures have been used to improve the processing efficiency of large-scale LLM services. However, the diversity of task requirements and the dynamics of resources pose great challenges to inference scheduling, leading to the wastage of many resources. In this paper, we present PerLLM, a personalized inference scheduling framework with edge-cloud collaboration designed for diverse LLM services. For the complexity of multiple constraints and the decision-making process of edge-cloud collaboration, we integrate the upper confidence bound algorithm based on the constraint satisfaction mechanism in PerLLM. For diverse LLM services, PerLLM can optimize service scheduling and resource allocation solutions within the edge-cloud infrastructure to meet processing time requirements while minimizing energy costs. Experimental results from different model deployments show that PerLLM can effectively meet the processing time requirements of personalized services. Compared to other methods, PerLLM achieves 2.2$\times$, 2.1$\times$, and 1.6$\times$ throughput and reduces the energy cost by more than 50\%.

\end{abstract}

\section{Introduction}
With the rapid development of artificial intelligence, large language models (LLMs) have attracted a lot of attention \cite{yang2024harnessing111}. Text generation has become increasingly popular in many applications, which has led to a surge in LLM services \cite{wang2024survey222}. For example, OpenAI's ChatGPT family currently has about 180 million users and receives over 1.6 billion service requests per month \cite{business333}. The main advantage of LLMs is their greater generative and learning ability. The massive volume of service requests generated by a large number of users requires real-time transmission and inference. Many LLMs are deployed on cloud servers for real-time inference \cite{zhang2019mark444}.

Although cloud servers offer significant computing and memory resources, processing all inference services in the cloud results in large-scale data transmission, which limits the performance of real-time inference due to limited bandwidth resources \cite{lin2023pushing555}. The high parameter model inference on cloud servers can lead to high energy costs \cite{patel2024characterizing666}. To reduce the number of model parameters, some researchers propose some model pruning \cite{ma2023llm777} and model compression \cite{zhu2023survey888} solutions. This makes it possible to deploy LLMs on edge servers \cite{dong2024creating999}. Edge computing can reduce the energy costs of LLM inference. But it can only support some simple tasks, service quality may be compromised when handling complex tasks.

\begin{figure*}[t!]
	\centering 
		\includegraphics[width=0.85\linewidth]{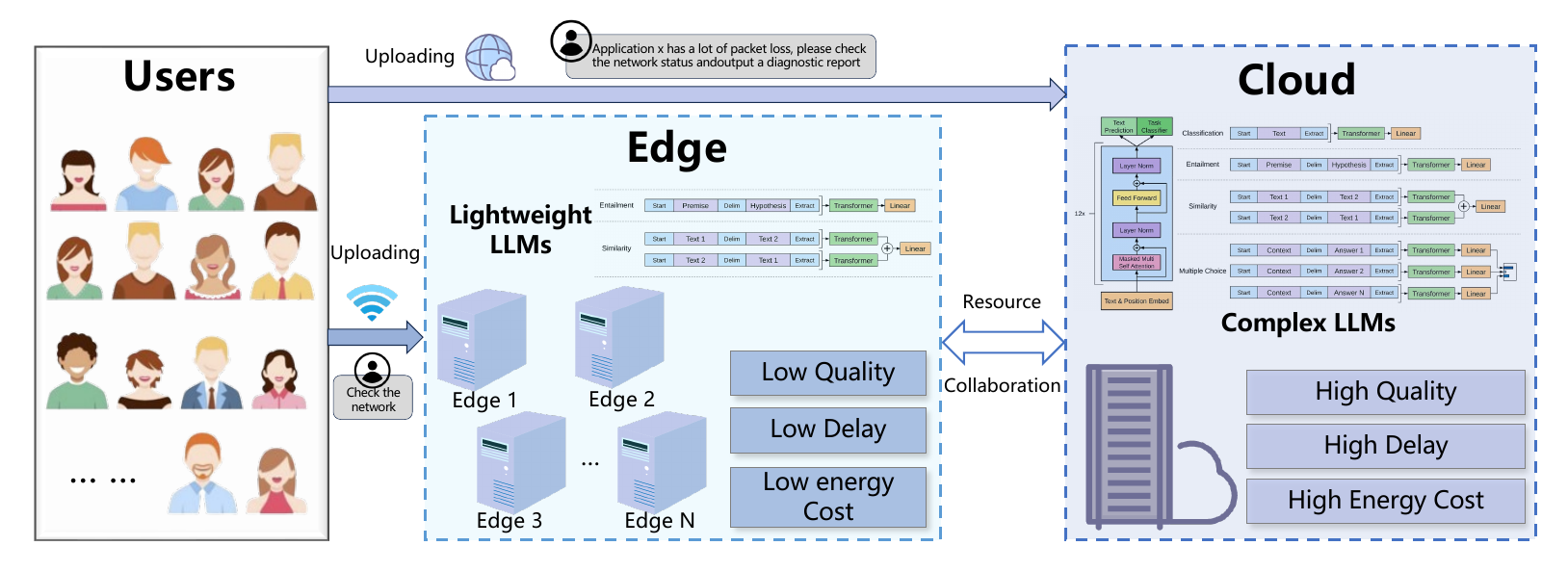}
	\caption{The illustration of edge-cloud collaborative inference architecture for LLM services.}
	\label{figure1}
\end{figure*}

In recent years, to solve the above problems, the edge-cloud collaboration architecture has emerged \cite{xu2024unleashing101010,fang2023large111111}, as shown in Figure~\ref{figure1}. It can utilize the advantages of both cloud computing and edge computing. Cloud servers usually have powerful computing and memory capabilities to provide high-quality inference services \cite{wang2023tabi121212}, but they also need a lot of energy costs. On the other hand, edge servers are closer to the user and can provide fast service responses while reducing energy costs \cite{yang2023visual131313}. It can effectively balance the demands for high-quality services and energy costs by distributing service requests to cloud servers and edge servers \cite{yang2024adaptive141414}. However, this architecture also presents some challenges. First, diverse services make task allocation in edge-cloud architectures difficult. For example, one user may need fast response time, while another user may be more concerned about the processing quality of long texts \cite{jin2024s151515}. In addition, the instability of network conditions and the dynamism of resources put high demands on the design of the scheduling system \cite{chen2023netgpt161616}.

To address the above issues, we propose PerLLM, a personalized inference scheduling framework that integrates edge-cloud collaboration. For diverse LLM services, PerLLM can optimize service scheduling and resource allocation solutions within edge-cloud infrastructures. The primary aim of this framework is to achieve adaptive scheduling of diverse LLM services under dynamic resource conditions and enhance the efficiency of LLM inference. The experiments demonstrate that our approach can significantly increase throughput and reduce processing time and energy costs.

The main contributions of this paper can be summarized as follows.

\begin{itemize}
\item We propose a novel personalized inference scheduling framework that leverages edge-cloud collaboration to handle massive inference services. It aims to maximize processing efficiency while satisfying the requirements for diverse LLM services.

\item We model the complex edge-cloud collaboration process as a combinatorial multi-armed bandit problem and propose a constraint satisfaction upper confidence bound algorithm for efficient service scheduling and resource allocation.

\item We evaluate the performance of PerLLM and compare it with baseline methods. Experimental results show that it can
achieve a more than 1.6$\times$ throughput and reduce the energy cost by more than 50\% while meeting the processing time requirements of diverse services.
\end{itemize}

\section{Preliminary and Motivation}

\subsection{Large Language Model}
LLMs represent a significant leap in the field of artificial intelligence, particularly within natural language processing (NLP). The core architecture behind most LLMs is the Transformer \cite{han2021transformer171717}, a type of deep learning model that uses the mechanism of self-attention \cite{vaswani2017attention232323} to weigh the influence of different words in a sentence. Self-attention mechanism is formulated as:

\begin{equation}
\text { Attention }(Q, K, V)=\operatorname{softmax}\left(\frac{Q K^T}{\sqrt{d_k}}\right) V,
\end{equation}

where $Q$ $K$, and $V$ denote the query, key, and value matrices, and $d_k$ is the dimension of the key. LLMs can generate coherent and contextually relevant text based on the input they receive. They can perform tasks such as text generation, translation, summarization, and sentiment analysis. The diversity of LLM services is a testament to their versatility, allowing for customization to meet the unique demands of different industries and applications \cite{borzunov2024distributed181818}. However, this diversity also introduces a high demand for resources. LLMs require substantial computing and memory resources to perform inference tasks \cite{oh2024exegpt191919}, which involve understanding and generating responses to user inputs. 

\subsection{Edge-Cloud Infrastructure}
Edge-cloud infrastructure represents a paradigm shift in the way computing resources are distributed and utilized \cite{villari2016osmotic202020}. It combines the strengths of both edge computing and cloud computing. At the core of edge-cloud infrastructure is the concept of bringing computation closer to the source of data generation. Edge computing \cite{shi2016edge373737} can reduce delay, energy costs, and bandwidth requirements since data does not need to travel long distances to reach a central data center. Cloud computing \cite{zhang2010cloud383838} can provide powerful computing and memory capabilities, making it ideal for more complex processing tasks and long-term data analytics. Moreover, edge-cloud infrastructure is crucial for deploying modern applications \cite{cao2021large212121}, especially for those that need to handle large-scale and diverse tasks, such as LLMs. Research on edge-cloud infrastructure focuses on optimizing computational offloading and resource allocation to achieve efficient inference of LLMs. 

\subsection{Observation and Motivation}
\begin{figure*}[t!]
	\centering 
    \subfloat[Cloud, Time ]{
    \label{Fig.sub.2.1}
    \includegraphics[width=0.23\textwidth]{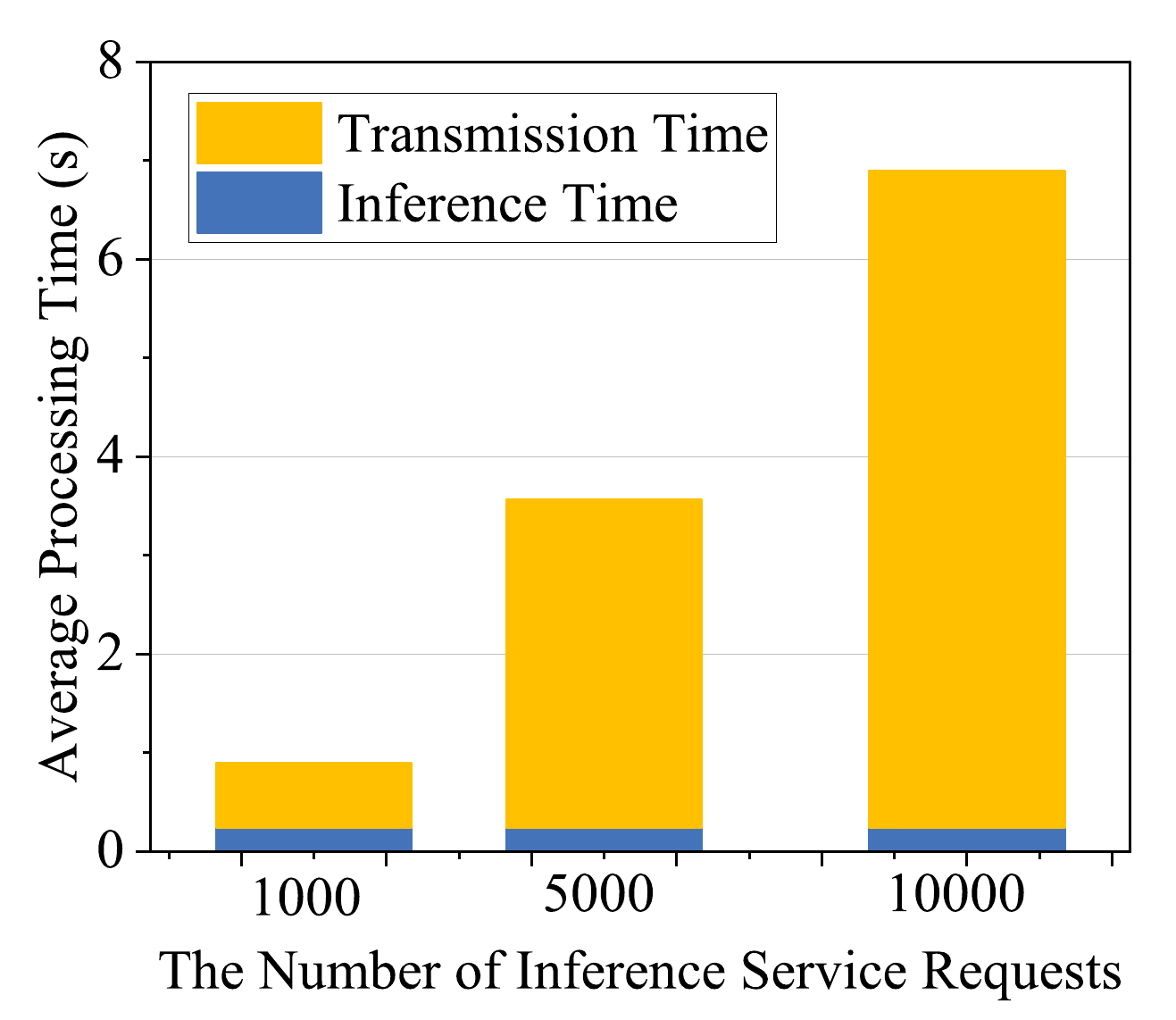}}
    \subfloat[Edge, Time]{
    \label{Fig.sub.2.2}
    \includegraphics[width=0.23\textwidth]{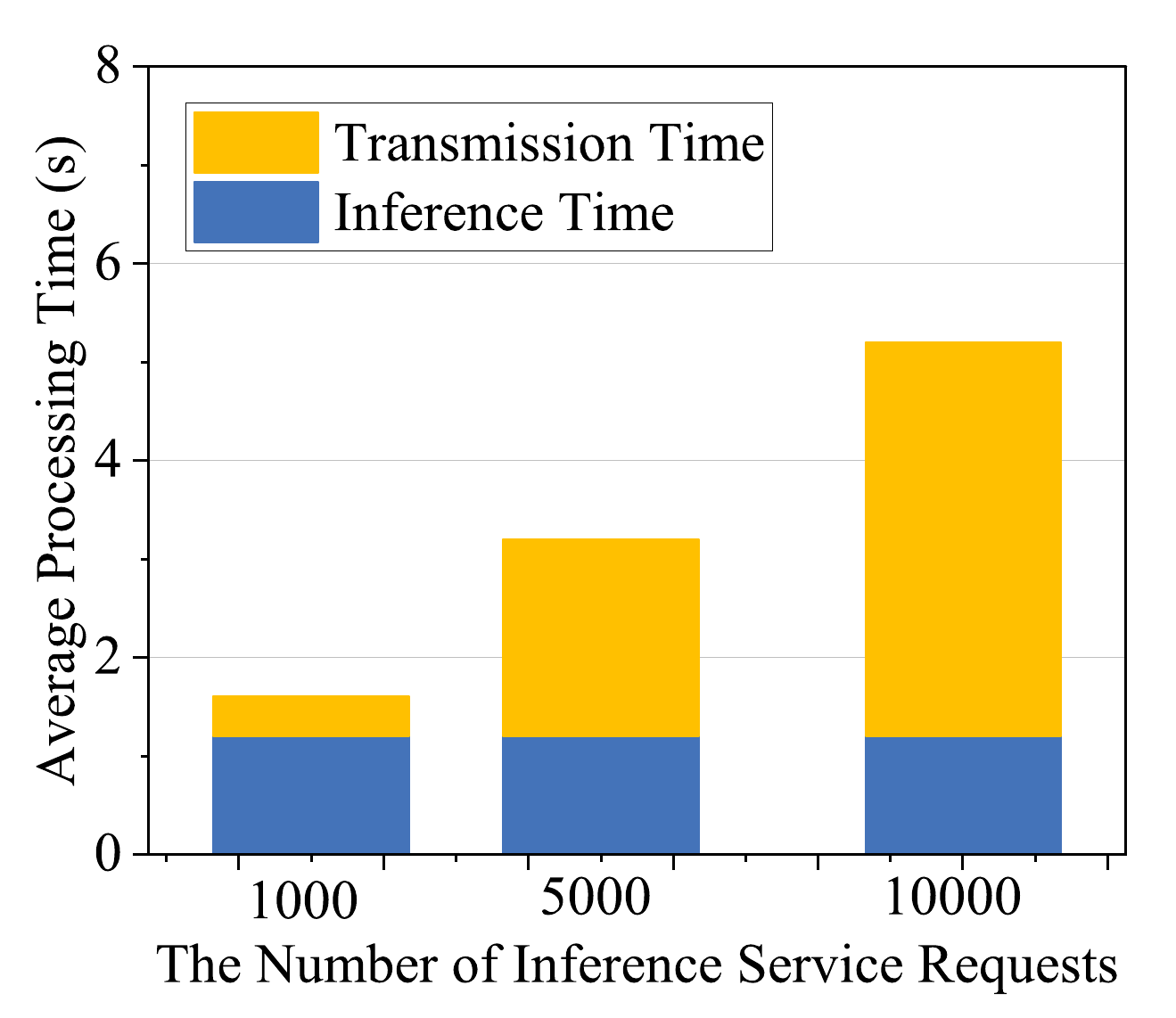}}
        \subfloat[Cloud, Energy Costs]{
    \label{Fig.sub.2.3}
    \includegraphics[width=0.24\textwidth]{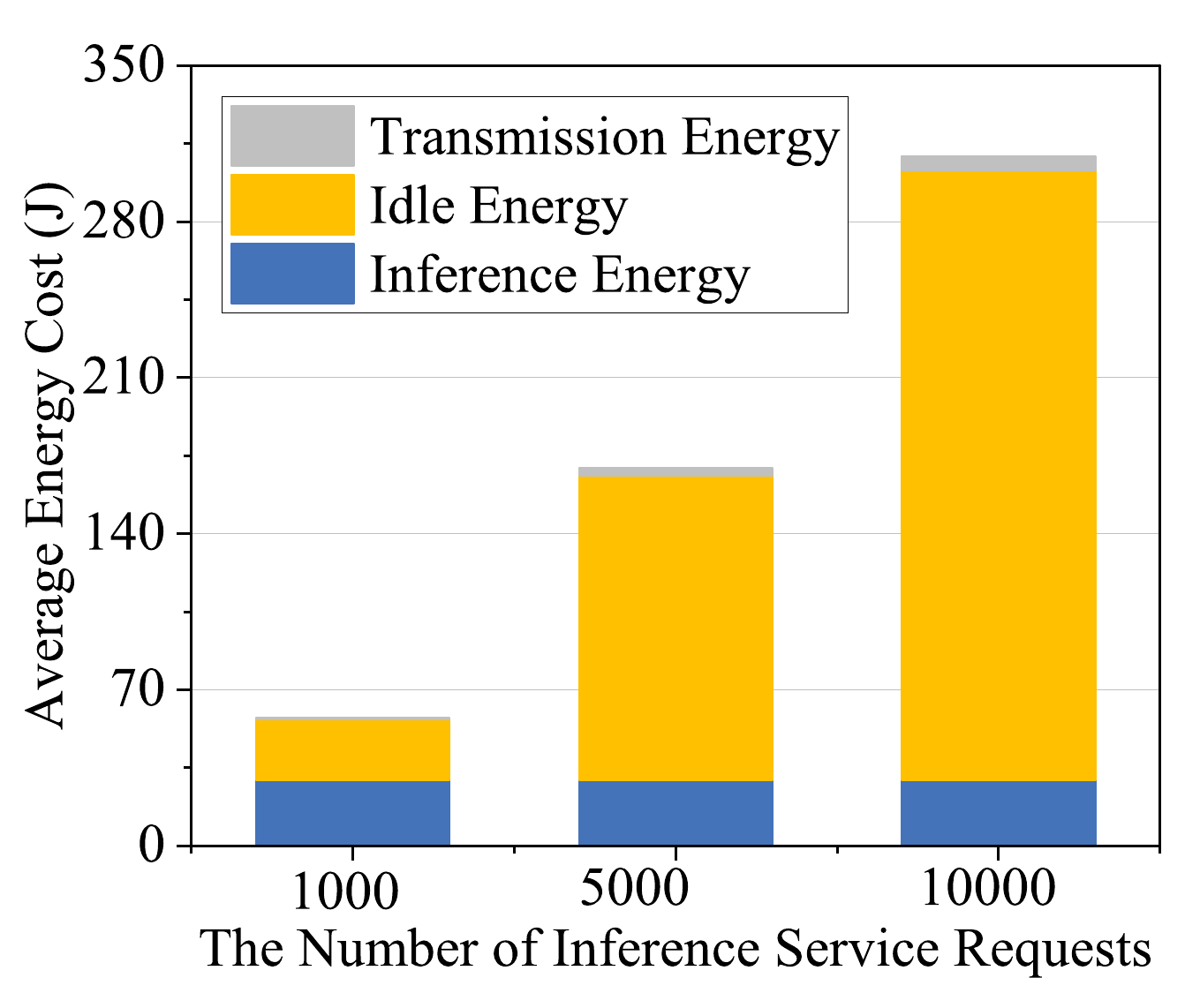}}
        \subfloat[Edge, Energy Costs]{
    \label{Fig.sub.2.4}
    \includegraphics[width=0.24\textwidth]{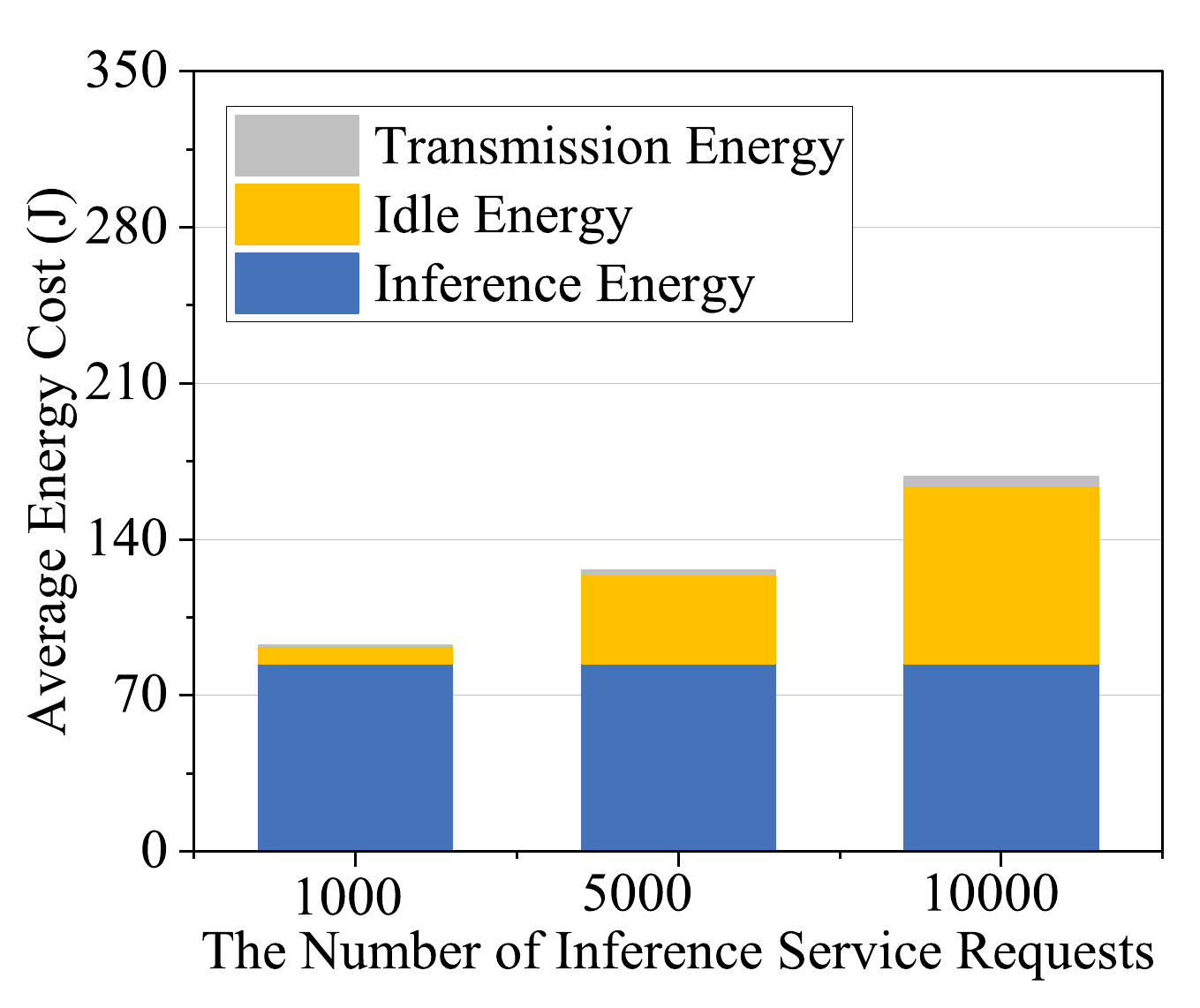}}
    \caption{The test results of average processing time and energy costs on cloud and edge.}
	\label{figure2}
\end{figure*}

We present results obtained from measurements conducted on actual devices, elucidating the necessity of adopting PerLLM. Five Intel Xeon Silver 4214R CPUs are used as edge servers, and one NVIDIA A100 GPU with 40 GB memory is used as the cloud server. 
We deploy the Llama-2-7B model on edge servers and the Llama-2-33B\footnote{\url{https://llama.meta.com/llama2/}}  model on cloud servers. The network bandwidth of the cloud and edge is set to 300 Mbps and 100 Mbps according to the work in \cite{yang2023javp222222}. The average test results per service are shown in Figure~\ref{figure2}, where processing time includes transmission time and inference time. The energy cost includes inference, transmission, and idle energy. 

We can find a surge in processing time and energy costs on the cloud when the number of inference services increases. This is because the simultaneous uploading of large-scale services causes network congestion. We also find the inference time of the edge server is longer than the cloud, and the transmission time is much shorter, leading to a significant advantage in total processing time over the cloud server. In addition, network congestion causes cloud servers to incur unnecessary energy costs, resulting in a much higher total energy cost than the edge server. Although energy cost savings differ by case, some energy cost reductions are usually achieved through edge-cloud collaboration. Therefore, we will investigate personalized inference scheduling and dynamic edge-cloud collaborative optimization solutions for diverse LLM services.


\section{PerLLM Design}

\begin{figure*}[t!]
	\centering 
		\includegraphics[width=1\linewidth]{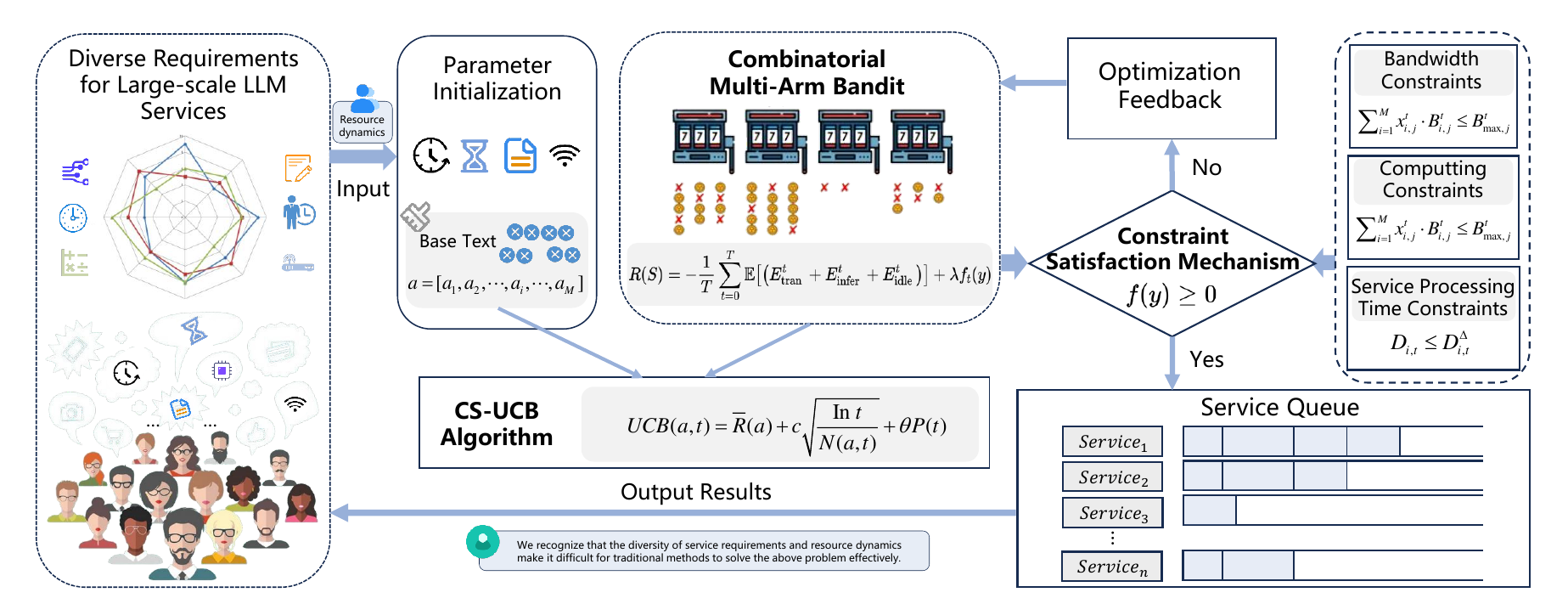}
	\caption{The overall design and workflow of the proposed PerLLM framework.}
	\label{figure3}
\end{figure*}

In this section, we present PerLLM, a personalized inference scheduling framework with edge-cloud collaboration. Figure~\ref{figure3} illustrates the overall design and workflow of PerLLM. Before using it, the lightweight LLM needs to be deployed at the edge and the complex LLM needs to be deployed in the cloud. Section 3.1 presents the problem formulation of PerLLM, detailing the optimization model and constraints. Then the solution algorithm employed is introduced in Section 3.2, and the theoretical analysis is described in Section 3.3.

\subsection{Problem Formulation}
Our focus is on the edge-cloud cooperative inference scheduling optimization problem for large-scale LLM services. We now define the problem statement more formally. We denote the set of input LLM services by  $G=\left\{g_1, g_2, \cdots g_i, \cdots g_M\right\}$ and  $U=\left\{u_1, u_2, \cdots u_j, \cdots u_N\right\}$ is used to represent the set of servers, where $s_N$ denotes the cloud server. These LLM services will be transmitted in real-time to the edge server or cloud server for inference. To guide how to balance service offloading and resource allocation between edge servers and cloud servers, a multi-objective optimization problem is formulated. The model minimizes the total energy cost, including transmission, inference, and idle energy, thereby reducing the overall energy costs while ensuring inference performance. Subsequently, we use processing time, bandwidth, and computing power as constraints. Processing time constraint is used to ensure that the service is completed in an acceptable time to the user, bandwidth and computing power constraints ensure compliance with current resource conditions. This can be formalized as follows: 

\begin{equation}
\begin{array}{ll}
 &  \min \quad \frac{1}{T} \sum_{t=0}^T\left(\omega_{\text{tran}} E_{\text {tran }}^t+\omega_{\text{infer}} E_{\text {infer }}^t+\omega_{\text{idle}} E_{\text {idle }}^t\right)\\

\text { s.t. } &  C_{1}: D_{i, t} \leq D_{i, t}^{\Delta}, i \in\{1,2, \cdots M\}, t \in T \\

& C_{2}: \sum_{i=1}^M x_{i, j}^t \cdot C_{i, j}^t \leq C_{\max , j}^t, \forall j, t \in T ,\\

& C_{3}: \sum_{i=1}^M x_{i, j}^t \cdot B_{i, j}^t \leq B_{\max , j}^t, \forall j, t \in T  \\

& C_{4}: \sum_{j=1}^N x_{i, j}^t=1, x_{i, j}^t=\{0,1\}, s_j \in S, t \in T

\end{array}
\end{equation}
where $E_{\text {tran }}^t$, $E_{\text {infer }}^t$, and $E_{\text {idle }}^t$ are the transmission energy, inference energy, and idle energy, respectively. $\omega$ is the weight factor. $D_{i, t}$ is the processing time for service $i$, and $D_{i, t}^{\Delta}$ is the processing time requirement. $x_{i, j}^t$ indicates whether service $u$ is assigned to the $j-$th server.  $B_{i, j}^t$ is the bandwidth requirement for service $i$ to upload to server $j$ and  $C_{i, j}^t$ is the computing power requirement. The constraint $C_{1}$ is used to ensure that the processing time requirement of each server is met. Otherwise, it is assigned to a more resource-rich server. Constraint $C_{2}$ indicates that the required computing power for all services is less than or equal to the currently available computing power. Constraint $C_{3}$ indicates that the bandwidth required to upload all services is less than or equal to the currently available bandwidths. Constraint $C_{4}$ ensures that only one server can be selected for each service in time slot $t$.
This formulation transcends the mere optimization of energy costs, which balances the reduction of energy costs with the requirements of performance. By considering bandwidth limitations, computing resource restrictions, and the stringent time requirements of inference services, the model adeptly navigates the complex environment of resource management.

\subsection{Solution Algorithm}
We recognize that the diversity of service requirements and resource dynamics make it difficult for traditional methods to solve the above problem effectively. Moreover, considering the complexity of multiple constraints leads to difficulties in satisfying these conditions simultaneously \cite{brakensiek2023sdps323232}. To address these difficulties, we conceptualize the above multi-objective optimization problem as a Combinatorial Multi-Armed Bandit (CMAB) problem \cite{chen2016combinatorial303030}. The CMAB problem reflects the multiple decision-making processes for service scheduling and resource allocation in edge-cloud infrastructures. We define the action space $a=\left[a_1, a_2, \cdots, a_i, \cdots, a_M\right]$ as assigning each service to a specific server. Where each element $a_i=\{1,2, \cdots j, \cdots, N\}$ denotes the index of the server to which the $i$-th task is assigned. The state space is $s=\left[\left(c_1, b_1\right),\left(c_2, b_2\right), \cdots,\left(c_j, b_j\right), \cdots,\left(c_N, b_N\right)\right]$, which represents the current computing and bandwidth resources of each server. To efficiently meet the diverse requirements of services and reduce the search space, we develop a constraint satisfaction mechanism. The scheme $y$ satisfies all constraints if $f(y) \geq 0$, otherwise it violates at least one constraint. $f(y)$ can be formulated as:

\begin{equation}
f(y)=\min \left(\frac{D_{i, t}^{\Delta}-\sum_{i=1}^M D_{i, t}}{D_{i, t}^{\Delta}}, \frac{C_{\max }^t-\sum_{i=1}^M C_i^t}{C_{\max }^t}, \frac{B_{\max }^t-\sum_{i=1}^M B_i^t}{B_{\max }^t}\right).
\end{equation}

To ensure that all considered solutions satisfy the given constraints, we integrate them into the optimization process. The constraint satisfaction upper confidence bound (CS-UCB) algorithm is proposed and the optimization process is shown in Algorithm 1.

\begin{algorithm}
\SetAlgoLined
 \textbf{Input}$:$\par
  \quad \,      The set of services $G$, the set of servers $Y$, the processing time requirements for services $ D_{i, t}$. \\

  \par
  
  \textbf{Output}$:$\par
 \quad \, The service and resource allocation scheme $X^*$, $S^*$.

 \textbf{Procedure}$:$ \\   
    {
\ \,1: \, Initialize parameters $\lambda$, $\alpha$, $\beta$ and $\delta$; \par 
\qquad Initialize the server's computing and bandwidth resources; \par \qquad Initialize the estimated rewards for all possible actions.\

\ \,2: \qquad \textbf{for} each search service\

\ \,3:  \qquad \quad \,  $  t \leftarrow t+1$

\ \,4:  \qquad \quad \, Play a super arm $S_t$

\ \,5: \qquad \quad \, \textbf{if} $D_{i,t} \leq D_{i, t}^{\Delta}$ and  $\sum_{i=1}^M C_i^t \leq C_{\max }^t$ and $\sum_{i=1}^M B_i^t \leq B_{\max }^t$  \textbf{then}

\ \,6: \quad \qquad \qquad   $  X^* \leftarrow x_{i,j}^{t}=1$,  
$  S^* \leftarrow S_{t}$ \ 

\ \,7: \qquad \quad \, \textbf{else} \ 

\ \,8: \quad \qquad \qquad Update $\operatorname{Reg}(T)$ and $S_{t}$ \ 

\ \,9: \qquad \quad \, \textbf{end if} \ 

10:\, \qquad \textbf{end for}\

11: \,  \textbf{return} $X^*$, $S^*$.

\par
    }
 \caption{Constraint Satisfaction Upper Confidence Bound (CS-UCB) Algorithm }
\end{algorithm}

In the CMAB problem, each super arm $S_t$ represents a resource allocation scheme. The reward obtained by playing a super arm is defined as follows:

\begin{equation}
R(S)=-\frac{1}{T} \sum_{t=0}^T \mathbb{E}\left[\left(\omega_{\text{tran}} E_{\text {tran }}^t+\omega_{\text{infer}} E_{\text {infer }}^t+\omega_{\text{idle}} E_{\text {idle }}^t\right)\right]+\lambda f_t(y).
\end{equation}

Specifically, we add the constraint satisfaction function to the reward. When there are constraints that are not satisfied, the reward is reduced. $\lambda$ is the coefficient of the constraint satisfaction function. To solve the challenge of attaining an exact solution, we introduce approximation coefficients $\alpha$ and $\beta$, where  $\alpha, \beta<1$. The introduction of $\alpha$ and $\beta$ allows for the delineation of an approximate regret function. They play a crucial role in steering the algorithm toward a feasible and approximate solution by measuring the acceptable level of approximation within the optimization process \cite{zuo2021combinatorial333333}. The approximate regret function is subsequently defined as:

\begin{equation}
\operatorname{Reg}(T)=T \cdot \alpha \beta \cdot R\left(S_{\max }\right)-\mathbb{E}\left[\sum_{t=1}^T R\left(S_t\right)\right],
\end{equation}

where $S_{\max }$ is the expected reward of the optimal super arm. The core of the algorithm is to compute an upper confidence bound for each possible action, which is mainly determined by the average reward $\bar{R}(a)$ and the number of choices $L(a, t)$ in the current action. It can be formulated as:
\begin{equation}
\operatorname{UCB}(a, t)=\bar{R}(a)+\delta \sqrt{\frac{\operatorname{In} t}{L(a, t)}}+\theta P(\mathrm{t}).
\end{equation}
The $\delta$ is a parameter used to control the balance between exploration and exploitation. In the decision-making phase, the CS-UCB algorithm first filters out all the actions that satisfy a given constraint through a constraint satisfaction mechanism and then selects the action with the highest UCB value $a_t=\arg \max _{a \in A} \operatorname{UCB}(a, t)$ among these valid actions for execution. As the algorithm iterates, the number of selections and estimated reward for each action are updated as a way to balance the relationship between exploring unknown actions and exploiting the known actions. It also updates the regret function. In this way, the algorithm can learn and optimize the decision-making process step by step while ensuring that the constraints are satisfied to find approximate solutions. Considering the changing service requirements, this approach is particularly suitable for scenarios that require dynamic decision-making under multiple constraints.

\subsection{Theoretical Analysis}
\textbf{Regret Bound Analysis}. In the CS-UCB algorithm, the performance bounds are usually expressed as bounds on cumulative regret. Cumulative regret is the sum of rewards lost in the case of choosing the suboptimal arm compared to the optimal choice \cite{carpentier2011upper343434}. With the introduction of the constraint satisfaction mechanism, the CS-UCB algorithm needs to consider constraint violations for each arm. We call the super-arm S a bad arm if $f(y) < 0$. We assume that there exists a penalty term $P(t)$ that represents the severity of $S$ as a bad arm.  The computation of regret bounds needs to take into account the additional cost introduced due to constraint violation. If $P(t)$ is designed to be proportional to the degree of violation, the regret bound can be estimated as:
\begin{equation}
\operatorname{Reg}(T) \leq \sqrt{2 M N \log (L)}+\theta P(\mathrm{t}),
\end{equation}
where $\theta$ is a conditioning parameter to balance the weights of reward and penalty. This shows that the regret grows logarithmically over time, suggesting that the CS-UCB algorithm can effectively learn and approximate the optimal policy.

\textbf{Complexity Analysis}. 
Suppose that there are $A$ possible actions, and constraint checking for each action requires a time complexity of $O(f(A))$. At each time step $t$, it is first necessary to compute the UCB values for each action and check whether the constraints are satisfied. The time complexity of this process is $O(A \cdot f(A))$. The time complexity of then updating the number of choices for each action and estimating the reward is $O(A)$. Thus, the overall time complexity of the algorithm is $O(A \cdot f(A) + A)$. In terms of space complexity, the algorithm needs to store the estimated gains and number of decisions for each action. Therefore the space complexity is mainly determined by the space required to store this information, which is $O(A\cdot L)$.

\section{Experimental Evaluation}

\subsection{Evaluation Setup}
\textbf{Implementation Details}. To verify the effectiveness of PerLLM, we use five Intel Xeon Silver 4214R CPUs as edge devices, and one NVIDIA A100 GPU with 40 GB memory is used as the cloud server. We simulate two network conditions with stable bandwidth settings of 100 Mbps and 300 Mbps for edge servers and the cloud server. The fluctuating bandwidth is set to vary within 20\% for simulating a dynamically changing environment. In addition, to simulate the simultaneous uploading of large-scale LLM services. We use Hugging Face's Transformers library\footnote{\url{https://github.com/huggingface/transformers}} to load the pre-trained model and the lexicon,  which generates text based on the input prompts. To ensure the diversity and quality of service generation, the temperature parameter is set to 0.8, and the top-k is set to 200.

\textbf{Models}. We use models ranging from 6 billion parameters to 33 billion parameters for evaluation.  Specifically, we deploy LLaMA2-33B on a cloud server. Then Yi-6B, LLaMA2-7B, LLaMA3-8B, and Yi-9B are deployed on edge servers, respectively.

\textbf{Evaluation Metrics}. We use processing time, throughput, and energy costs as the main performance metrics. Processing time includes transmission time and inference time. Throughput is the number of tokens processed by the system per unit time. Energy costs include transmission energy, inference energy, and server idle energy.

\textbf{Baseline Methods}. We compare our solution with the following three baseline methods. (1) FineInfer\cite{he2024deferred313131}: This is a cloud-only solution that utilizes delayed sequential batch processing for task scheduling. (2) AGOD\cite{du2024diffusion242424}: This is an edge-only solution for service offloading through diffusion modeling and deep reinforcement learning. (3) RewardlessGuidance\cite{fang2023large111111}: This is an edge-cloud solution that optimizes offloading decisions and resource allocation using a rewardless guidance algorithm.

\subsection{Processing Time Analysis}
\begin{figure*}[t!]
	\centering 
    \subfloat[Stable Bandwidths]{
    \label{Fig.sub.4.1}
    \includegraphics[width=0.45\textwidth]{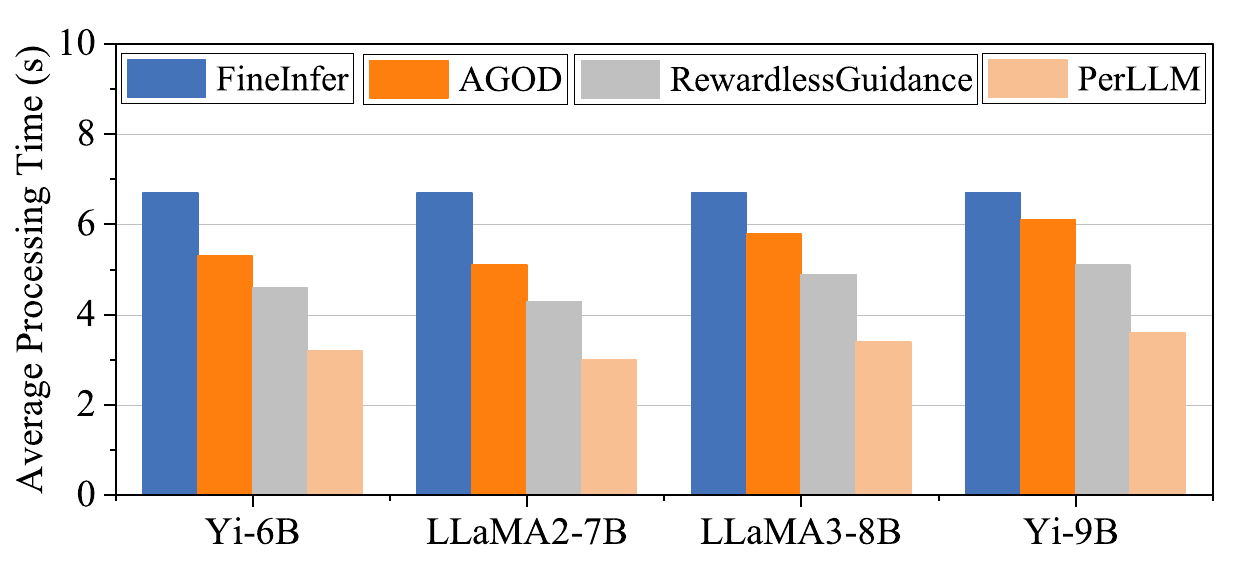}}
    \subfloat[Fluctuating Bandwidths]{
    \label{Fig.sub.4.2}
    \includegraphics[width=0.45\textwidth]{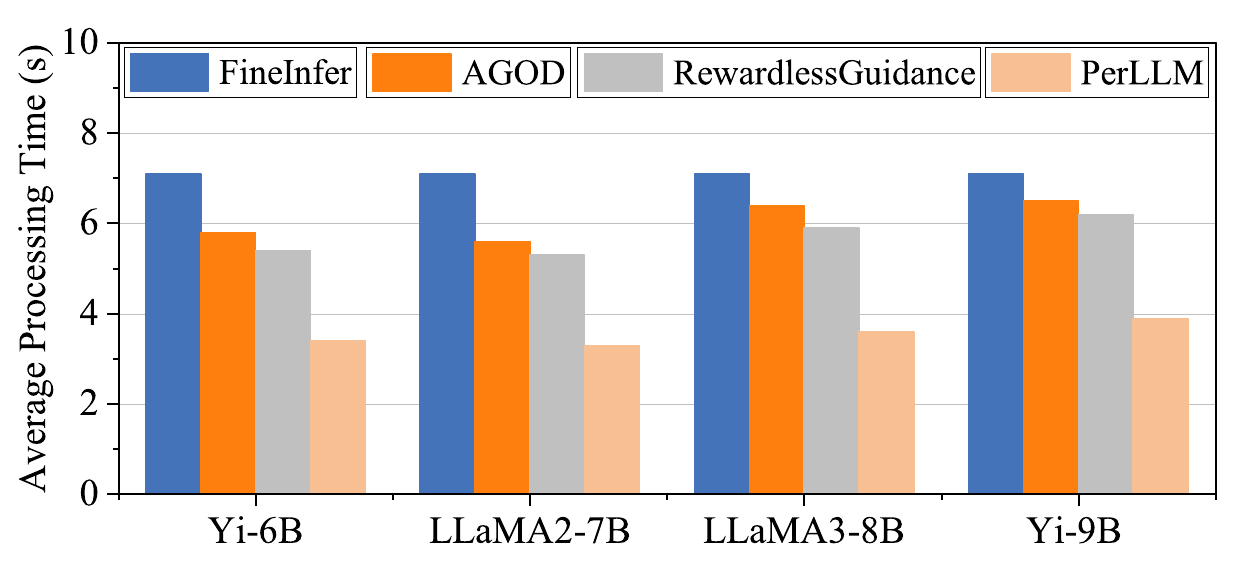}}
    \caption{The average processing time comparison of different methods under different models.}
	\label{figure4}
\end{figure*}

We first evaluate the average success rate of meeting the service's predefined processing time requirements in different bandwidth environments. Specifically, a service is judged successful if its final processing time is less than the input time requirement. Conversely, tasks that fail to meet these criteria are categorized as unsuccessful. To model the diversity of service requirements, we randomly select the processing time requirements of the input services from [2s, 6s], which represent a wide range of application requirements \cite{gan2023model353535}. To simulate LLM services with high concurrency, the number of inference service requests is set to 10000. The comparison results under different model deployments are shown in Table~\ref{tab1}. The results show that PerLLM has a success rate of over 97\%. Notably, even with fluctuating bandwidths, PerLLM maintains a high success rate. Other methods are only 58\%-74\%. We then show the comparison results of the average processing time per service under the different methods in Figure~\ref{figure4}. It can be found that for different model deployments, the processing time of PerLLM is lower than other methods. The advantage in bandwidth fluctuating environments is even more obvious. PerLLM can achieve a higher success rate and lower processing time because the introduced constraint satisfaction mechanism can dynamically allocate resources according to service requirements.

\begin{table}[th]
  \caption{The average success rates for meeting the processing time requirements of services.}
  \label{tab1}
  \centering
  \tabcolsep 7pt 
  \begin{tabular}{ccccccccl}
    \toprule
  Different Models &FineInfer & AGOD & RewardlessGuidance & PerLLM     \\\hline
  Yi-6B & 58\%  & 67\% & 74\%  &  \quad \textbf{98\%}  \\
  LLaMA2-7B &  58\% & 69\% & 77\%  & \quad  \textbf{99\%} \\
  LLaMA3-8B&  58\% & 66\% & 74\%  & \quad  \textbf{98\%} \\
  Yi-9B &  58\% & 66\% & 71\%  & \quad  \textbf{97\%} \\
  \bottomrule
\end{tabular}
\end{table}

\subsection{Throughput Analysis}
\begin{figure*}[t!]
	\centering 
    \subfloat[Stable Bandwidths]{
    \label{Fig.sub.5.1}
    \includegraphics[width=0.45\textwidth]{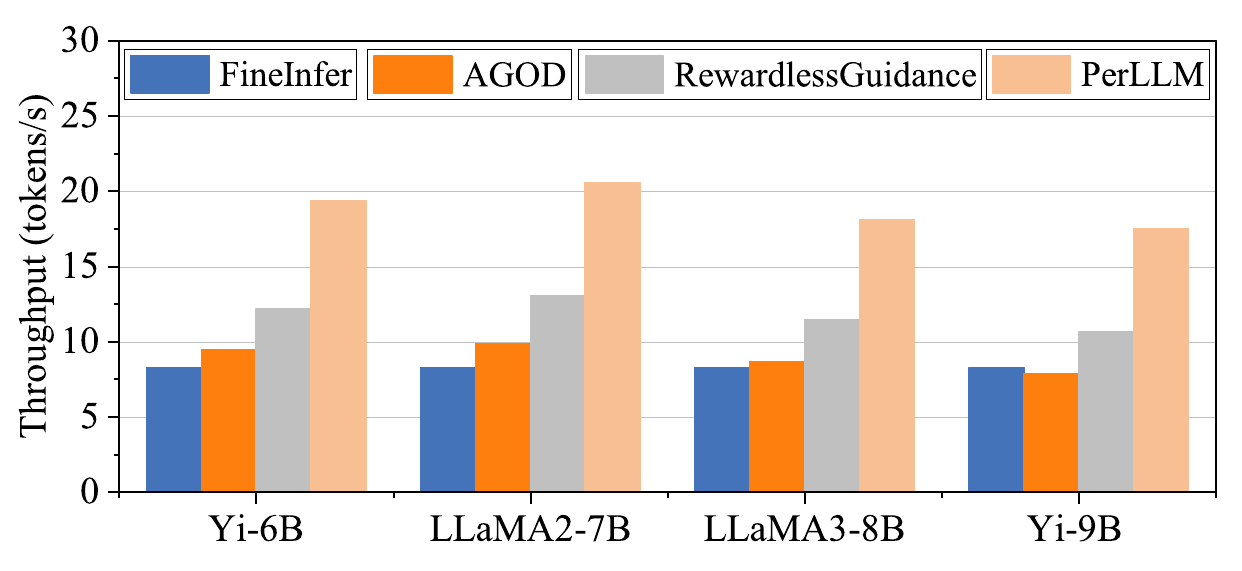}}
    \subfloat[Fluctuating Bandwidths]{
    \label{Fig.sub.5.2}
    \includegraphics[width=0.45\textwidth]{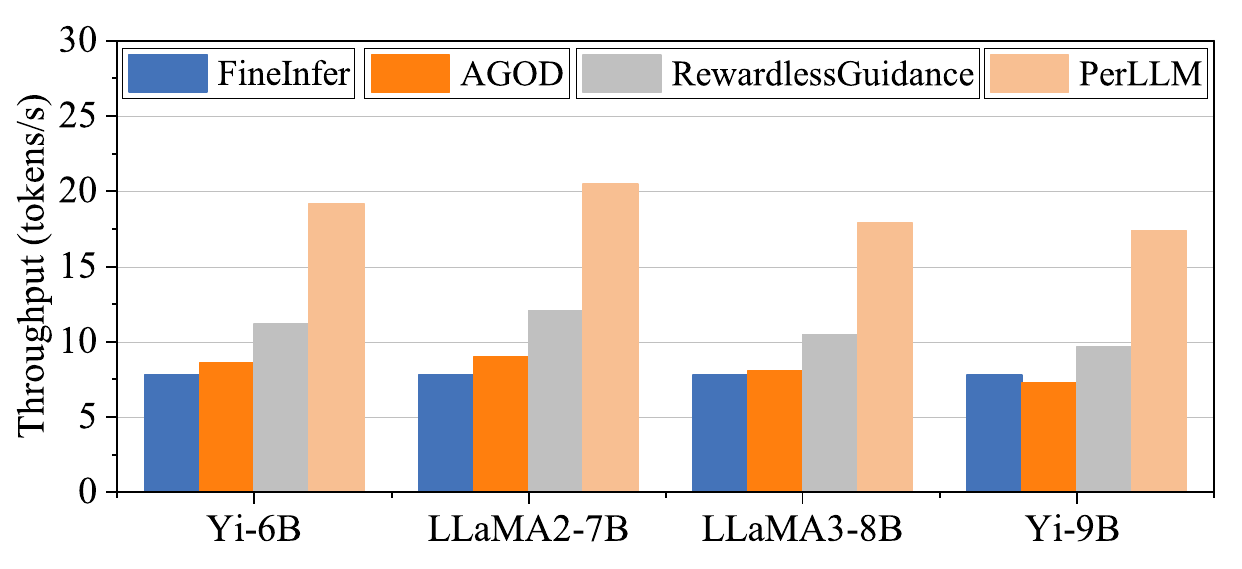}}
    \caption{The throughput comparison of different methods under different models.}
	\label{figure5}
\end{figure*}

Throughput is a crucial metric for evaluating the performance of LLM services and edge-cloud infrastructures \cite{sheng2023flexgen363636}, as it directly impacts the system's ability to handle large volumes of requests efficiently. In this context, we conduct a comprehensive evaluation of the throughput of the proposed PerLLM. Figure~\ref{figure5} shows the comparison results of throughput under different approaches. The results show that PerLLM significantly outperforms other methods regarding throughput for both stable and fluctuating bandwidths. The cloud-only solution FineInfer has the lowest throughput due to network bandwidth limitations. The edge-only solution AGOD also has poor throughput results due to computational resource constraints. The RewardlessGuidance method improves some performance but still suffers from poor scheduling. On average, PerLLM achieves 2.2$\times$, 2.1$\times$, and 1.6$\times$ throughput when comparing FineInfer, AGOD, and RewardlessGuidance. This indicates that our proposed solution can avoid network congestion and queuing problems through rational service scheduling and resource allocation. 

\subsection{Energy Cost Analysis}
\begin{figure*}[th]
	\centering 
    \subfloat[Stable Bandwidths]{
    \label{Fig.sub.6.1}
    \includegraphics[width=0.45\textwidth]{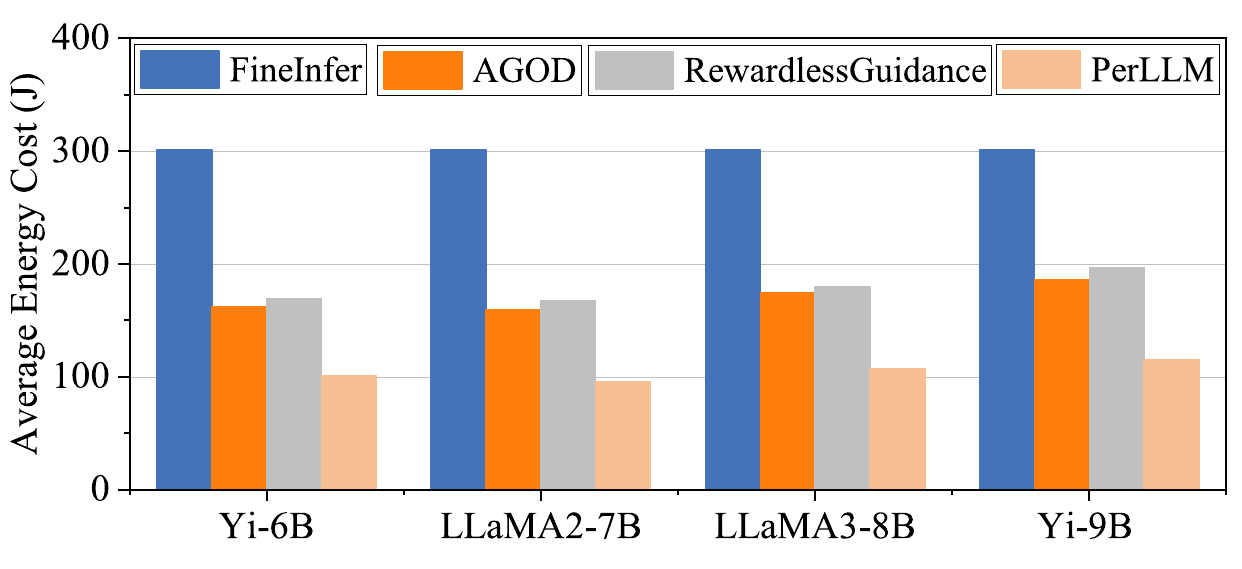}}
    \subfloat[Fluctuating Bandwidths]{
    \label{Fig.sub.6.2}
    \includegraphics[width=0.45\textwidth]{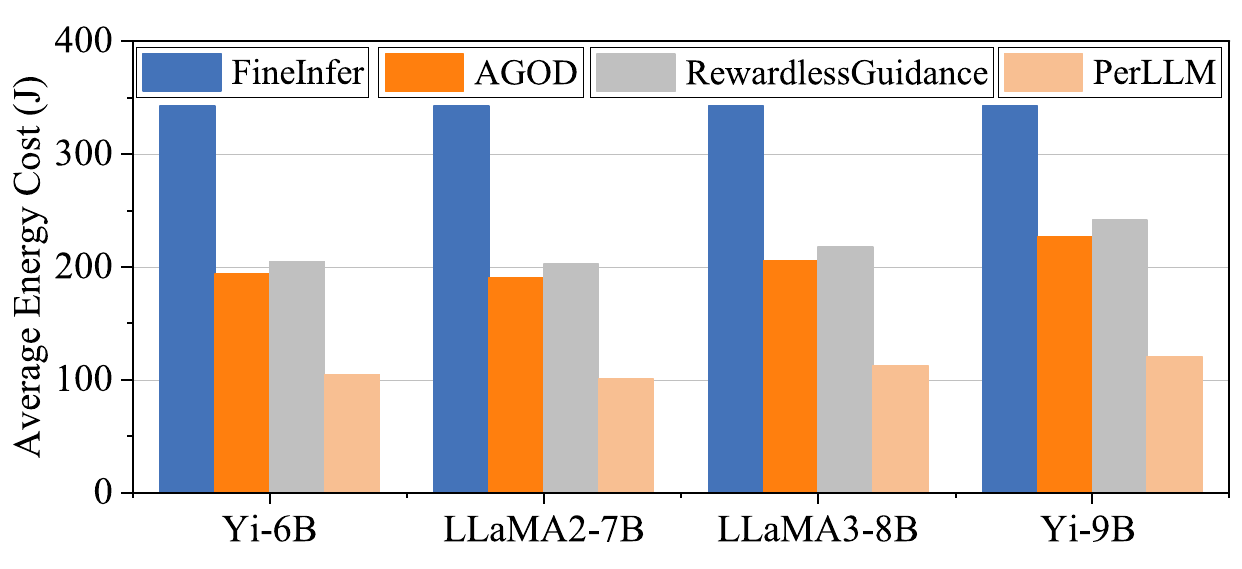}}
    \caption{The energy cost comparison of different methods under different models.}
	\label{figure6}
\end{figure*}

Finally, we evaluate the energy cost of the different methods. 
Server inference energy refers to the energy consumed during the processing of service requests, while server idle energy accounts for the energy used when the server is on standby but not processing. Transmission energy costs involve the energy required to transfer data between servers and users. Figure~\ref{figure6} illustrates the comparison results for energy costs under different methods. PerLLM adapts well to dynamic scenes, consistently demonstrating lower energy costs than other methods. The power of cloud services is usually high, so the FineInfer method incurs more energy costs. The other two methods also consume more energy cost. This is because the diversity of service requirements and dynamic resources make it difficult for them to schedule loads efficiently, thus incurring many unnecessary energy costs. Overall, PerLLM can reduce energy costs by more than 50\%. It mitigates the environmental impact associated with high energy costs in LLM services. This strategy towards personalized services ensures that PerLLM can be adapted to more application scenarios.

\section{Related Work}
\textbf{LLM Service Offloading with Edge-Cloud Infrastructure.} The high interest in LLM has sparked numerous research in service offloading optimization. For large-scale LLM inference services,  the authors in \cite{fang2023large111111} propose a proactive inference method based on a reward-free bootstrap. It uses the expected future free energy to optimize the offloading decision to solve the LLM inference service offloading problem in edge-cloud systems. To enable LLM service providers to provide better user experiences, the authors in \cite{du2024diffusion242424} improve the efficiency of service offloading by combining the optimal decision algorithm based on the diffusion model with deep reinforcement learning. The authors in \cite{li2024multi252525} propose an attention-enhanced multi-intelligent body reinforcement learning algorithm to support generative model-driven edge learning. It improves the efficiency of executing LLM services on edge servers through a multi-task computational offloading model. The authors in \cite{wang2024toward262626} propose a framework to reduce the inference cost of LLM services by transferring subtasks to mobile edge networks. They then introduce a deep reinforcement learning algorithm to optimize the selection of edge servers. In real-world scenarios, LLM service requirements are constantly changing due to diverse users. The above methods ignore the analysis of service requirements and make it difficult to realize efficient edge-cloud collaborative service offloading.

\par
\textbf{Resource Allocation with Edge-Cloud Infrastructure.} 
Simultaneous uploads of large-scale LLM services lead to resource constraints \cite{dhar2024empirical393939}. Some researchers focus on resource allocation to improve the inference efficiency of LLM services. The authors in \cite{oh2024exegpt191919} present ExeGPT, a distributed system designed for constraint-aware LLM inference. To adapt to the workloads of different nodes, they propose a scheduling strategy based on cyclic allocation and workload awareness. It can maximize inference throughput while satisfying a given delay constraint. The authors in \cite{hu2024laecips272727} design an edge-cloud collaboration strategy based on hard input mining and optimization of resource allocation. They propose to update the edge model and its collaboration strategy with the cloud under the supervision of the LLM. For the high computational and memory costs required for LLM services\cite{kwon2023efficient404040}, the authors in \cite{liu2024spa282828} propose a side plugin adaption method based on feature knowledge. It establishes interaction between a pre-trained LLM on the cloud and additional parameters on the edge. The work in \cite{wang2023privatelora292929} allocates privacy-sensitive computations to edge servers and common computations to the cloud. It effectively mitigates bandwidth resources by leveraging the low rank of residual activations. While these efforts have achieved some progress, they continue to face notable limitations. These methods often rely on specific scenarios and static states, lacking adaptability to highly dynamic network and computing environments. Moreover, they often do not account for real-time fluctuations in workload and the various requirements of users, which can drastically affect performance.


\section{Discussion}

\textbf{Advantages Analysis}. The PerLLM offers several notable advantages in addressing the rapid growth of LLM users and the limitations of bandwidth-constrained cloud servers. First, by leveraging edge-cloud collaboration, PerLLM significantly enhances the processing efficiency of large-scale LLM services. Second, PerLLM can accommodate the diverse requirements of LLM services by integrating a constraint satisfaction mechanism and providing a personalized inference scheduling approach. Third,  the CS-UCB algorithm allows for effective handling of the complexity arising from multiple constraints and dynamic decision-making processes. By optimizing service scheduling and resource allocation within the edge-cloud infrastructure, PerLLM can minimize energy costs and promote sustainable development.

\textbf{Limitations Analysis}. PerLLM in the current version has some limitations. First, despite its high success rate of over 97\% in dynamic environments, there are still some services that fail to meet processing time requirements. The reason behind these failures could be the limitations of the resource allocation algorithm or the unforeseen complexity of the service itself. Second, the PerLLM framework does not yet support accuracy and memory optimization. Moreover, the same equipment is used for multiple edge servers, and the heterogeneous edges are not yet considered. This field is still in its infancy, in the future, we will endeavor to address the above limitations.


\section{Conclusion and Future Work}
In this paper, we introduce PerLLM, a personalized inference scheduling framework that leverages edge-cloud collaboration to enhance the processing efficiency of LLM services. Considering the multiple constraints and the complexity of edge-cloud collaboration, we propose a CS-UCB algorithm based on the constraint satisfaction mechanism to ensure optimal decision-making under conditions of uncertainty. The main goal of this framework is to achieve adaptive scheduling of diverse LLM services under dynamic resource conditions. Compared with state-of-the-art solutions, experimental results show that PerLLM can meet the processing time requirements of more than 97\% of services. It also achieves more than 1.6$\times$ throughput and reduces the energy cost by more than 50\%.

In the future, we will explore multi-dimensional resource collaborative optimization in edge-cloud infrastructures for LLM services, with a particular focus on memory integration. Additionally, exploring how to improve accuracy through edge-cloud collaboration and continuous learning mechanisms is also an interesting study for the future.

\small
\bibliography{cite}

\end{document}